# Kinetic electrocaloric effect and giant net cooling of lead-free ferroelectric refrigerants

Yang Bai[1,2], Guangping Zheng[1] and Sanqiang Shi[1]

[1]Department of Mechanical Engineering, Hong Kong Polytechnic University, Hung Hom, Kowloon, Hong Kong
[2]Key Laboratory of Environmental Fracture (Ministry of Education), University of Science and Technology Beijing, Beijing 100083, China
Email: baiy@mater.ustb.edu.cn and mmzheng@polyu.edu.hk

**Abstract.** Giant electrocaloric (EC) effect is observed in $BaTiO_3$ multilayer thick film structure. The temperature change is as high as 4.0 °C under an applied electric field of 352 kV/cm. Most importantly, the EC effect is found to depend on the varying rate of the applied field. Based on the giant net cooling (~0.37 J/g) resulting from the difference in the varying rates of rising and falling fields, the kinetic EC effect provides an effective solution for the design of refrigeration cycle in ferroelectric micro-refrigerator.

# 1. Introduction

Solid state refrigeration based on ferroelectric refrigerants has attracted considerable attention in recent years because of its high reversible adiabatic temperature change, good efficiency and feasible miniaturization. In particular, the ferroelectric micro-refrigerator is the best solution for the cooling of microelectronic devices and microelectromechanical systems [1]. The ferroelectric refrigeration is based on the electrocaloric (EC) effect, which is referred to the heat exchange resulting from the electric field induced paraelectric-ferroelectric (P-F) phase transition. Recently, the giant EC effect is obtained in some ferroelectric thin films successively [1-6], and was studied in various ferroelectric materials based on thermodynamics [7-13]. The large EC reversible adiabatic temperature change (> 10 °C) of those ferroelectric thin films makes them most desirable to be used as refrigerants in ferroelectric micro-refrigerators. In spite of encouraging progresses, there are several critical issues that have hindered the practical application of those ferroelectric thin films as refrigerants in ferroelectric micro-refrigerators. First, the heat absorption capacity of a thin film is too small for practical application in ferroelectric refrigeration due to the tiny volume of the working medium. Second, how to effectively implement the refrigeration cycle of the ferroelectric refrigerants, i.e, how to design heat exchanges among the ferroelectric refrigerants, heat sink and heat sources, is still an open question which always exists in solid state refrigeration. For example, although Ericsson cycle has been well identified as the best refrigeration cycle for ferroelectric refrigeration, complex and expensive heat switches or mechanical shifters [14-16] have to be used, which in practice is difficult to be implemented in micro-refrigerator.

We solve the above-mentioned issues by developing the BaTiO$_3$ multilayer thick film structure (MLTFS) as ferroelectric refrigerants. Considering that the giant EC effect of thin films originates from the ultrahigh applied electric field (hundreds of kV/cm), dozens times higher than the breakdown field of bulk ceramics, in this study, alternatively, the ferroelectric thick films with multilayer structure is selected as the refrigerant for ferroelectric micro-refrigerator, which has the advantages of both ultrahigh breakdown electric field and large volume of working medium. Our previous work [17] and Kar-Narayan's report [18] both demonstrated the high EC effect in MLTFS under ultrahigh electric field.

In this paper, the giant EC effect of these MLTFSs is investigated by direct measurement using differential scanning calorimeter (DSC) and we focus on the kinetics of heat exchanges during the EC processes. Typically, the P-F phase transition is a first-order phase transition (FOPT) and its energy conversion involved is always determined by the kinetic process [19], i.e., how the applied electric field is varied, but the investigation on the kinetics of EC effect is still lacking up to date. In our work, a giant net cooling in one EC cycle is realized by kinetics control without the aid of any other accessory. The refrigeration cycle can be easily designed based on such kinetic EC effect.

**2. Experimental**

The MLTFS was fabricated by the tape-casting method. The sample had a multilayered structure similar to that of a multilayer chip capacitor. The ferroelectric medium of the $BaTiO_3$ (EYANG Technology Development Co., Ltd.) and the inner Ni electrode were printed alternately and cofired. In the structure, two groups of interpenetrating electrodes led to two terminals, respectively. The average thickness of the $BaTiO_3$ layers was 1.4 μm and the total number of dielectric layers was 180.

The direct measurement of the heat flow was conducted as an isothermal process using a DSC (TA Instruments Q200), and a DC power supply (Agilent N8741) was used to apply the electric field to the sample. The sample was connected to the DC power by two vanished wires, which were too thin (d~0.01mm) to change the thermal isolation of the testing cell. The sample was attached to the DSC pan with insulating glue. In the experiment, the start and end electric field intensities were fixed to 0kV/cm and 176kV/cm, respectively. The rising rate of the driving field varied from 1.4kVcm$^{-1}$s$^{-1}$ to 176kVcm$^{-1}$s$^{-1}$, and the falling rate was kept constant at 176kVcm$^{-1}$s$^{-1}$.

## 3. Results and discussion

Fig. 1 (a) and (b) show the DSC measurements of the MLTFS sample at 40$^o$C and 80$^o$C respectively, where the rising field rate varies from 1.4 kVcm$^{-1}$s$^{-1}$ to 176 kVcm$^{-1}$s$^{-1}$ and the falling field rate is kept at 176 kVcm$^{-1}$s$^{-1}$. There are obvious exothermal and endothermal peaks corresponding to the application and removal of the electric field, respectively. The EC endothermal values is 0.70 J/g at 40$^o$C and 0.89 J/g at 80$^o$C under a falling field rate of 176 kVcm$^{-1}$s$^{-1}$, which are much stronger than that in most reported ferroelectric ceramics, either lead-free or lead-based. The corresponding temperature change *ΔT* are 1.3 K and 1.8 K respectively. Under higher electric field, the EC effect can be further improved. A heat absorption of 2.02 J/g (*ΔT*

= 4.0 K) is obtained at 80$^{\circ}$C under 352 kV/cm. The giant EC effect in MLTFS is caused by the ultrahigh electric field, which is much higher than the applied field in ceramics (~30 kV/cm). During the EC process, phase transition is induced by an external electric field, the work of which is converted into lattice elastic energy and polarization energy through lattice deformation and the generation and reorientation of polarization. Hence, the higher the field intensity, the higher the latent heat of the phase transition. In addition, it is clearly shown in Fig. 1 that both the exothermal and endothermal heat flows return to the baseline very well after enough thermal equilibrium time, demonstrating the negligible Joule heating during EC process. The large EC temperature change of MLTFS samples suggests that they are suitable and practical for ferroelectric refrigeration application.

It is clear from Fig.1 that the exothermal value decreases as the rising field rate drops, whereas the endothermal value remains constant under a fixed falling field rate. The difference in the endothermal and exothermal values in an electric field cycle suggests that the EC effect is determined not only by static parameters, such as electric field intensity and temperature, but also by kinetics processes, such as the rising and falling rates of electric field. Net cooling in a single EC cycle can be realized by controlling the driving field rate. Table 1 lists the heat values of the exothermal and endothermal processes under different driving field rates. The net heat adsorption value in an electric field cycle is as high as 0.37 J/g, which is even higher than the EC heat absorption value (the area of the endothermal peak) reported for most lead-based ferroelectric ceramics.

Text pages

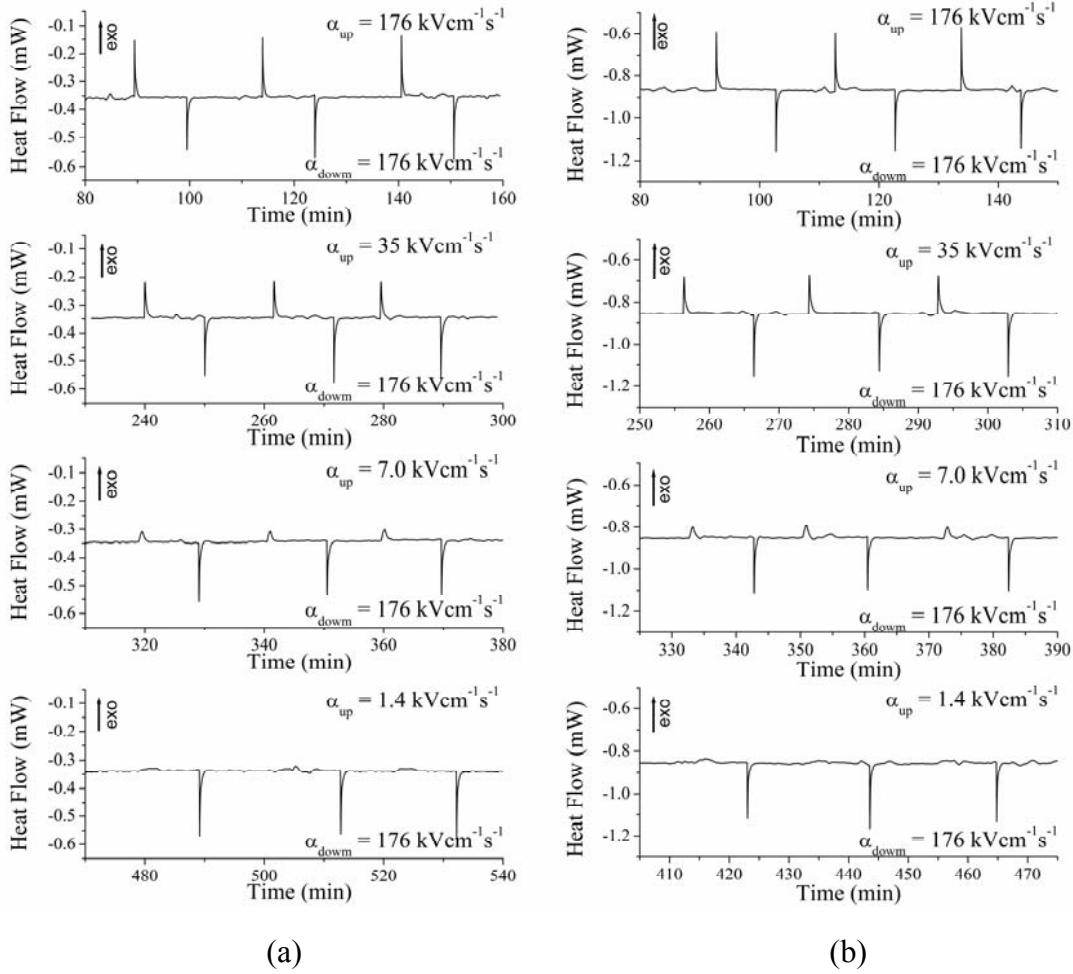

(a)                                              (b)

Fig. 1: The DSC heat flow of the sample under 176kV/cm at (a) 40°C and (b) 80°C. The sample is a parallel connection of eight BaTiO$_3$ MLTFSs. $\alpha_{up}$ and $\alpha_{down}$ are linear varying rates of the rising and falling electric fields respectively.

The relation between the varying rate of electric field and the heat value of exothermal or endothermal process can be fitted using a general power-law relation,

$$L = L_0 + A\alpha^n \qquad (1)$$

where $\alpha$ is the varying rate of the linear driving field, **n** is the scaling exponent, $L_0$ is the value of heat exchange in a quasistatic case ($\alpha \sim 0$), and *A* is a constant. Fig. 2 plots the linear relation between log(**L-L$_0$**) and log$\alpha$, and shows them to have a good linear correlation of $R_{40}$ = 0.95 and $R_{80}$ = 0.98 at working temperatures of 40°C and 80°C, respectively. The scaling exponents **n** for these two temperatures are both 0.135.

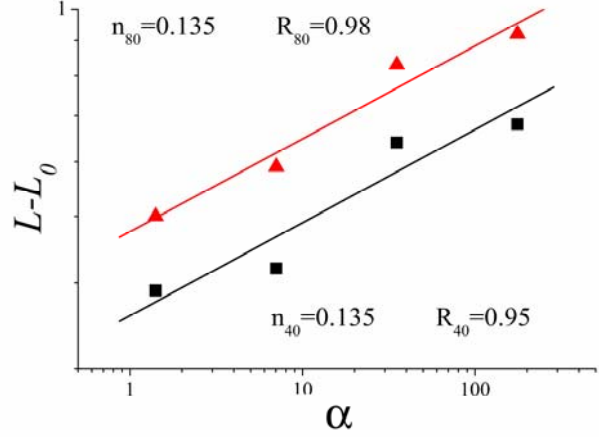

Fig. 2: Double logarithmic graph of (*L–L$_0$*) and driving electric field rate *α*.

Table 1: The measured exothermal values (*H$_{exo}$*) and endothermal values (*H$_{endo}$*) under different driving field rates.

|  | $\alpha_{up}$ (kVcm$^{-1}$s$^{-1}$) | $H_{exo}$ (J/g) | $\alpha_{down}$ (kVcm$^{-1}$s$^{-1}$) | $H_{endo}$ (J/g) |
|---|---|---|---|---|
| 40°C | 176 | 0.69±0.01 | 176 | 0.70±0.03 |
|  | 35 | 0.64±0.03 | 176 | 0.70±0.02 |
|  | 7.0 | 0.42±0.02 | 176 | 0.68±0.03 |
|  | 1.4 | 0.38±0.05 | 176 | 0.69±0.04 |
| 80°C | 176 | 0.92±0.03 | 176 | 0.89±0.02 |
|  | 35 | 0.83±0.03 | 176 | 0.87±0.05 |
|  | 7.0 | 0.59±0.03 | 176 | 0.89±0.04 |
|  | 1.4 | 0.50±0.09 | 176 | 0.87±0.05 |

The ferroelectric phase transition is a typical FOPT related to nucleation and growth kinetics. We further explain the kinetics of the electrocaloric effect by using the Landau theory of ferroelectrics. Under the time-dependent rising or falling electric field, the free energy of the materials is expressed as

$$F = \frac{1}{2}aP^2 + \frac{1}{4}bP^4 + \frac{1}{6}cP^6 - P \cdot E(t) \qquad (2)$$

where *P* is the polarization, *a* = *a$_0$*(*T*-*T$_c$*)/*T$_c$*, and *T$_c$* is the Curie temperature. *a$_0$*, *b* and *c* are coefficients independent of temperature. We assume *E*(*t*) varies linearly with time *t* with a varying rate of *α*, as it was in our experiment. Because of the nonequilibrium relaxation of the polarization during FOPT [19-21], its kinetics can be described by the Ginzburg-Landau equation as follows,

$$\frac{dP}{dt} = -\frac{\delta F}{\delta P} = -aP - bP^3 - cP^5 + E(t). \quad (3)$$

From Eq. (2), the entropy can be expressed as

$$S = -\frac{\partial F}{\partial T} = -\frac{a_0}{2T_c} P^2. \quad (4)$$

The endothermal or exothermal heat resulting from the removal or application of an electric field can be calculated by

$$\Delta Q = T\int_{S_0}^{S_1} dS = -\frac{a_0}{T_c} T \int_{E_0}^{E_1} P(\frac{dP}{dE})dE = -\frac{a_0}{T_c} T \int_0^{\Gamma} P(\frac{dP}{dt})dt, \quad (5)$$

where $E_0$ and $E_1$ are the applied fields at time $t=0$ and $\Gamma$ respectively, and corresponding entropies of the materials are $S_0$ and $S_1$ respectively. By solving Eq. (3), the relation between heat conversion $\Delta Q$ and varying rate $\alpha$ of the electric field is found to follow a power law when $E_0$ and $E_1$ are fixed:

$$(\Delta Q - \Delta Q_0) \propto \alpha^{2/3}, \quad (6)$$

where $\Delta Q_0$ is an α independent heat conversion during static F-P transition. Although the exponent $n=2/3$ is different with that of $n=0.135$ in Eq. (1), Ginzburg-Landau theory predicts the same scaling relation between the endothermal or exothermal heat and $\alpha$ as that (Eq. (2)) observed in experiment

The scaling coefficient $n=0.135$ is also much smaller than the value of 1/2 obtained from a phenomenological ferroelectric phase transition models which considers the effect of impurities on the energy conversion [22]. The difference could be caused by the significant lattice deformation and the resulted changes of polarization under ultrahigh electric field applied on $BaTiO_3$ thick films, which can not be well described by either Landau theory or phenomenological model.

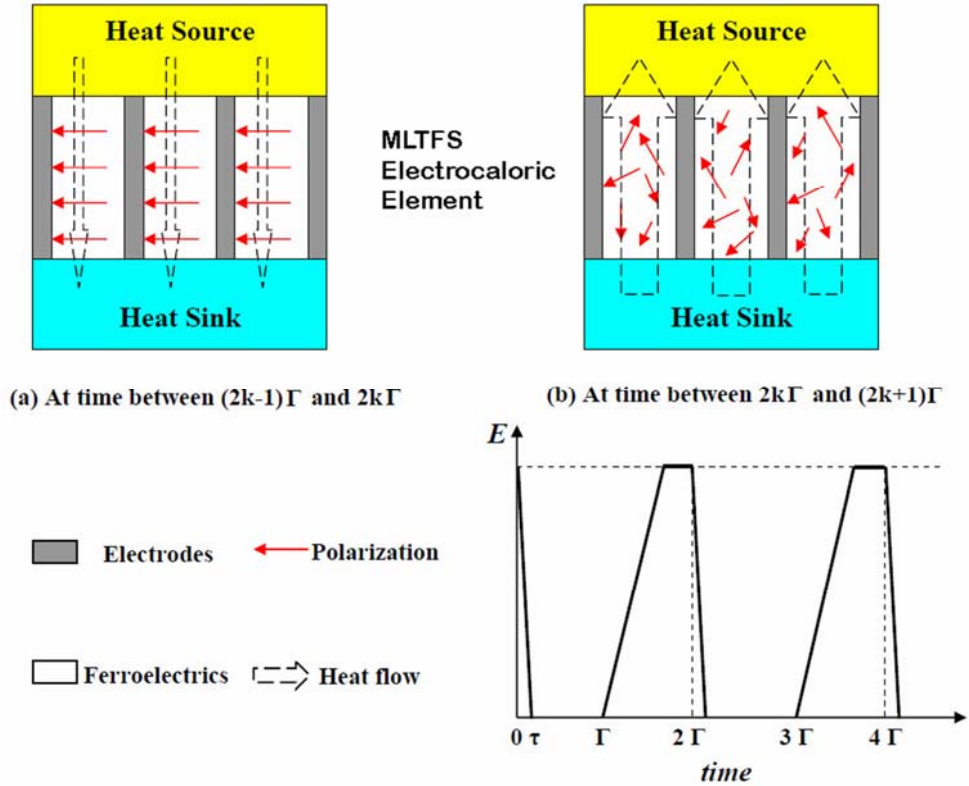

Fig. 3: Schematic of $BaTiO_3$ MLTFS used as refrigerants. Net cooling in one electric field cycle (2 Γ) is achieved: heat dissipation of MLTFS element at time interval 0 to Γ (a) is smaller than heat adsorption at time interval Γ to 2 Γ (b). *k* is an integer.

   The kinetics studies of the EC effect are very important for the practical application of EC refrigeration. Because ferroelectric refrigeration is based on a solid-state phase transition, the ferroelectric material acts as both a warm junction and a cold junction in different sections of the refrigeration cycle. In previous design [14-16, 23] of refrigeration cycle, a micromechanical shifter or heat switch was necessary to control the direction of heat flow. The mechanical shifters work by opening and closing the thermal contact between the cooler and the object. The heat switches made from semiconductor or liquid crystal are operated by varying the heat conductance under different conditions [14-16]. Obviously, those designs have complex structure, high cost, low reliability and poor robustness for micro-refrigerators. Instead, from our work, a feasible alternative method of refrigeration is provided using a simple ferroelectric MLTFS based on the scaling relation between EC heat and the driving field rate. Giant net cooling (~0.37 J/g) in an electric field cycle can be achieved by controlling the driving electric field rates. A schematic of the design based on the kinetic EC effect of the MLTFS ferroelectric refrigerant is shown in Figure 3. Because the rate of rising electric field is smaller than that of the falling field, there is a net cooling in an electric field cycle (time interval $\Delta t=2\Gamma$). Such design for ferroelectric refrigeration can significantly reduce the accessory used to accomplish a full refrigeration cycle. For example, it is not necessary to use micromechanical shifter or heat switch since the ferroelectric MLTFS elements can acts as heat switches as well.

## 4. Conclusions

In summary, we demonstrate the giant EC effect of $BaTiO_3$ multilayer thick film structure by direct DSC measurement and characterize the scaling relation between the EC heat and the driving field rate. The MLTFS samples show a giant electrocaloric effect with heat absorption values of 0.70 J/g at $40^oC$ and 0.89 J/g at $80^oC$ under an electric field of 176 kV/cm. The measured EC heat is much higher than that of most reported ferroelectric ceramics because of the ultrahigh applied electric field. The exothermal and endothermal values during the application and withdrawal of electric field are respectively determined by the rising and falling rate of electric field, and the driving field rate dependence of heat value follows a general power-law relation. A giant net cooling (~0.37 J/g) in one electric field cycle can be obtained by applying different rising and falling rates of electric field. This work provides a feasible method to implement net cooling in one electric field cycle without the aid of any other accessory, which will lead to a practical application of lead-free ferroelectric micro-refrigerator in near future.

**Acknowledgement**

This work was supported by a grant from the Research Grants Council of the Hong Kong Special Administrative Region, China (Project No. PolyU 7195/07E), the Research Funds of Hong Kong Polytechnic University (Project Nos. A-SA29, G-YX0X), and the National Science Foundation of China under grants of 50702005. G.P.Z. is grateful for the supports provided by the Innovation and Technology Commissioner of the Hong Kong Special Administrative Region, China (Project No. ITS/314/09).